# Calculation of impurity density and electron-spin relaxation times in p-type GaAs:Mn

Veronika Burobina[a)]

*Ioffe Institute, Politekhnicheskaya Street 26, Saint Petersburg 194021, Russia*

Magnetic semiconductors have aroused interest due to their various functionalities related to spintronic devices. Manganese (Mn) as a substitutional impurity in $A_3B_5$ semiconductors supplies not only holes, but also localized spins. The ejection of Mn atoms with an uncompensated magnetic moment leads to the appearance of ferromagnetic properties. The most suitable material characterized by long-term spin dynamics is n-type GaAs. In p-type GaAs, the spin relaxation time of electrons is generally much shorter. For purposes of this research, electron-spin relaxation times in 3D and 2D p-type GaAs were studied. Calculation of impurity densities and charge state of magnetic acceptors demonstrate the essential composition of the material. Comparison of theoretical and experimental data in optical-spin orientation of electrons reveal the longest spin relaxation time of 77 ns in 2D GaAs:Mn, less than twice the best time in the p-type 3D GaAs material.

---

[a)] E-mail: nikabour@mail.ioffe.ru

**Introduction**

The development of functional materials with enhanced properties[1,2,3] is a prominent subject of modern studies of science and technology. Spin phenomena in semiconductor materials is of considerable interest due to the use of the spin degree of freedom for recording, storing, and reading information.

The material p-type GaAs has been actively studied from different perspectives[4,5,6], and GaAs:Mn has been considered for a long time as a prominent material for spintronics applications[7,8,9]. The current examination of the p-type GaAs is related to determination of the composition of the material with the experimentally-determined prolonged electron-spin relaxation time. This method involves the analytical calculation of the concentrations of deep and shallow impurity concentrations $N_{Mn}$ and $N_d$ and $N_a$ respectively, using the charge neutrality condition for a partially-compensated acceptor semiconductor, as well as the consequential fitting of a plot of the dependences of spin relaxation times on excitation power for 3D and 2D p-type GaAs based on the experimental measurements of the long-spin relaxation time in the 3D material. The result reveals the best long-spin relaxation time of 77 ns for the quantum well, which is less than twice the atypically long-spin relaxation time in the 3D p-type GaAs:Mn[10].

**Calculation of impurity concentrations in GaAs:Mn**

Mn-doped GaAs is a magnetic, partially-compensated acceptor-type semiconductor. To evaluate the composition of the material, we started with calculations of deep impurities of manganese as well as shallow impurities of donors and acceptors based on available concentrations of holes in the valence bands obtained from Hall measurements at $T_1 = 300K$ and $T_2 = 77K$. The calculation of impurity concentrations were performed as follows.

The charge neutrality condition for a partially-compensated acceptor semiconductor of GaAs:Mn is

$$N_{Mn}^- + N_a^- = p + N_d, \quad (1)$$

where $N_{Mn}^-$ and $N_a^-$ are the concentrations of the ionized deep and shallow acceptor impurities, $p$ is the hole concentration in the valence band, and $N_d$ is the donor concentration. $N_d$ is equivalent to $N_d^+$ because GaAs:Mn is a partially-compensated acceptor semiconductor and, therefore, does not have neutral donor atoms. The condition (1) can be written in an alternative fashion, having determined $N_{Mn}^-$ and $N_a^-$ via the distribution functions and density of impurity states. The position of the Fermi level can be determined via the hole concentration in the valence band for a nondegenerated semiconductor. In this case, the charge neutrality condition will be

$$\frac{N_{Mn}}{1+\left(\frac{g_1}{g_0}\right)_{Mn} \times \frac{p(T)}{N_V(T) \times e^{\left(-\frac{E_{Mn}}{kT}\right)}}} + \frac{N_a}{1+\left(\frac{g_1}{g_0}\right)_{Mn} \times \frac{p(T)}{N_V(T) \times e^{\left(-\frac{E_a}{kT}\right)}}} = p(T) + N_d, \quad (2)$$

where $N_V(T)$ is the effective density of states in the valence band, $g_0$ and $g_1$ are the degeneration factors of an empty acceptor impurity center of Mn (occupied with an electron) and filled acceptor impurity center of Mn (occupied with a hole), $E_{Mn}$ and $E_a$ are the activation energies of deep and shallow acceptors, respectively, and $k$ is the Boltzmann constant.

Knowing the value of $E_a$ ($E_a$= 25 - 30 meV), the charge neutrality condition can be simplified. The comparison of the experimental hole concentrations (Table 1) obtained from the Hall measurements in the valence band at $T_1 = 300K$ and $T_2 = 77K$, and the hole concentrations derived from the expression $p(T) = N_V(T) \times e^{\left(-\frac{F-E_V}{kT}\right)}$ demonstrate that $p(T) \ll$ $N_V(T) \times e^{\left(-\frac{E_a}{kT}\right)}$. Thus, the charge neutrality condition for the partially-compensated semiconductor of GaAs:Mn can be rewritten as

$$\frac{N_{Mn}}{1+\left(\frac{g_1}{g_0}\right)_{Mn} \times \frac{p(T)}{N_V(T) \times e^{\left(-\frac{E_{Mn}}{kT}\right)}}} = p(T) + N_d - N_a \tag{3}$$

The system of equations for temperatures $T_1$ and $T_2$ from the condition (3) will be

$$\frac{N_{Mn}}{1+\left(\frac{g_1(T_1)}{g_0}\right)_{Mn} \times \frac{p(T_1)}{N_V(T_1) \times e^{\left(-\frac{E_{Mn}}{kT_1}\right)}}} = p(T_1) + N_d - N_a$$

and (4)

$$\frac{N_{Mn}}{1+\left(\frac{g_1(T_2)}{g_0}\right)_{Mn} \times \frac{p(T_2)}{N_V(T_2) \times e^{\left(-\frac{E_{Mn}}{kT_2}\right)}}} = p(T_2) + N_d - N_a$$

The solution of this system allows one to calculate the concentrations of $N_{Mn}$ and $N_d - N_a$. The degeneration factor of an empty Mn center is $g_0 = 6$, since $g_0 = 2s + 1$, where $s$ is the spin of a Mn ion, $= \frac{5}{2}$, and in which condition the 3d valence subshell of a Mn ion has five electrons. $g_1(T_1)$ and $g_1(T_2)$ were calculated from $g_1(T) = \beta_1 + \sum_{F=2}^{4} \beta_F \times e^{\left(-\frac{E_F}{kT}\right)}$, where $\beta_1$ is the ground state degeneracy of a Mn center, $\beta_F$ is the F-excited state degeneracy, $(\beta_F = 2F + 1)$, and $E_F$ is the energy of excited states relative to the ground state $(E_2 = 4.4\ meV, E_3 = 11\ meV, E_4 = 19.8\ meV)$. The values of these energies were calculated in the following way. The Hamiltonian of the exchange interaction of Mn ion electrons and a hole bound with the ion is $\widehat{H} = I\hat{s}\hat{j}$[11], where $I$ is the interaction constant and $j$ is the spin of a hole, $j = \frac{3}{2}$. The Hamiltonian $\widehat{H}$ can be written as $\widehat{H} = \frac{I}{2}[(\hat{s} + \hat{j})^2 - \hat{s}^2 - \hat{j}^2]$. The total moment of a Mn center filled with a hole is $\hat{F} = \hat{s} + \hat{j}$, under which condition $\hat{H} = \frac{I}{2}[\hat{F}^2 - \hat{s}^2 - \hat{j}^2] = \frac{I}{2}[F(F+1) - s(s+1) - j(j+1)]$. The total eigenvalues are determined from $s - j$ to $s + j$. Hence, $F = 1, 2, 3, 4$. It is known that the energy difference between the first excited state, $F = 2$, and the ground state, $F = 1$, is $\Delta E_{2-1} = 4.4\ meV$. The exchange interaction constant, therefore, can be calculated from $\Delta E_{2-1} = \frac{I}{2}[2(2+1) - 1(1+1)] = 2I$. Thus, $I = 2.2\ meV$. The energy difference between the second excited state, $F = 3$, and the first excited states will be $\Delta E_{3-2} = 6.6\ meV$. The energy difference between the third excited state, $F = 4$, and the second excited state is $\Delta E_{4-3} = 8.8\ meV$. Thus, the energies of the excited states relative to the ground state will be:

$$E_2 = 4.4\ meV\ \text{for the first excited state,}$$

$$E_3 = 11\ meV\ \text{for the second excited state, and}$$

$$E_4 = 19.8\ meV\ \text{for the third excited state.}$$

The degeneration factors of Mn filled centers, therefore, are $g_1(T_1) = 15.98$ and $g_1(T_2) = 7.37$.

The solution of the system of equations (4) will be

$$N_{Mn} = \frac{p(T_2) - p(T_1)}{\left[1 + \frac{g(T_2)}{g_0} \times \frac{p(T_2)}{N_V(T_2) \times e^{\left(-\frac{E_{Mn}}{kT_2}\right)}}\right] - \left[1 + \frac{g(T_1)}{g_0} \times \frac{p(T_1)}{N_V(T_1) \times e^{\left(-\frac{E_{Mn}}{kT_1}\right)}}\right]}$$

and

$$N_d - N_a = \frac{p(T_2)\left[1 + \frac{g(T_2)}{g_0} \times \frac{p(T_2)}{N_V(T_2) \times e^{\left(-\frac{E_{Mn}}{kT_2}\right)}}\right] - p(T_1)\left[1 + \frac{g(T_1)}{g_0} \times \frac{p(T_1)}{N_V(T_1) \times e^{\left(-\frac{E_{Mn}}{kT_1}\right)}}\right]}{\frac{g(T_2)}{g_0} \times \frac{p(T_2)}{N_V(T_2) \times e^{\left(-\frac{E_{Mn}}{kT_2}\right)}} - \frac{g(T_1)}{g_0} \times \frac{p(T_1)}{N_V(T_1) \times e^{\left(-\frac{E_{Mn}}{kT_1}\right)}}}$$

The calculated concentrations of deep $N_{Mn}$ and shallow $N_d - N_a$ acceptor impurities in different samples of GaAs:Mn are shown in Table 1.

**Discussion**

In p-type GaAs, the spin relaxation time does not depend on the pumping power if donor concentration is smaller than the shallow acceptor concentration Ref. [7]. To obtain long spin relaxation time in the p-type GaAs:Mn with a manganese deep concentration of approximately $10^{17} cm^{-3}$, the shallow donor concentration should be larger than shallow acceptor concentration. Then, the atoms of shallow acceptors and Mn deep impurity will be totally and partially ionized, respectively, at 8K. According to the experiment on optical orientation of electrons[12], under the influence of light, the photoexcited electrons, after being captured by donor impurities from the conductivity band, recombine with manganese ions generating the charge center of Mn. The effective magnetic field produced by these Mn centers influences the spins of electrons of the superior energy levels. Under the recombination of a hole from the valence band with a Mn ion, the spins of the hole and the Mn ion produce the antiferromagnetic state. Therefore, the total effective magnetic field of the Mn atom will be compensated and will not influence electron spin relaxation in the p-type GaAs:Mn. Thus, for further consideration, it is worthwhile to consider samples 5, sample 12, sample 21, and sample 22 of GaAs:Mn from Table 1 to obtain long-spin relaxation times in p-type GaAs.

**Spin relaxation time in 2D p-type GaAs:Mn**

We used the system of equations for the movement of charge carriers to facilitate the dependence of electron spin relaxation time vs. excitation power for a 2D p-type GaAs and compared this result with the earlier-calculated dependence of the electron-spin relaxation time vs. the excitation power, Ref. [7] for 3D p-type GaAs:Mn with a manganese concentration of $7.8 \times 10^{17} cm^{-3}$.

In a 2D quantum well of Mn-doped GaAs, electron spin relaxation time will be twice shorter than in the 3D material as it is described by the derived Equation (12) in Ref. [7]. The ratio of $\frac{N_d - N_a}{N_{Mn}}$, however, is $7.12 \times 10^{-2}$ for the quantum well of GaAs:Mn. Fig.1 shows the plotted dependences of electron-spin relaxation times on the excitation power for 2D and 3D p-type Mn-doped GaAs. The maximum-reached spin relaxation times for the 2D and 3D material are $77\ ns$ and $154\ ns$, respectively. The reduction of electron-spin relaxation time in the quantum well in comparison with 3D GaAs:Mn is due to the decrease of manganese concentration and the rate of hole recombination from the valence band to manganese levels. The ratio of the rate of hole recombination from the valence band to manganese levels to the rate of electron recombination from the conduction band to manganese levels for the quantum well is almost four times less than for the bulk material. Thus, the effective magnetic field produced by the spins of the manganese ions will be larger for the quantum well. The enhancement of the spin relaxation rate in the quantum well compared to the bulk, according to Dyakonov-Perel mechanism[13] is dominant contribution and is equivalent to $\frac{4}{(k_F \times L)^4} = 11$, where $k_F$ is the Fermi vector for the 3D density, and $L = 100$ Å — the well width.

The applied magnetic field can polarize the Mn spins, which alters the spin relaxation due to the s-d exchange scattering mechanism[14]. Hence, the expression for electron spin relaxation time in 2D p-type GaAs:Mn in the applied magnetic field, parallel to the quantum well plane along the x-axis, can be written as

$$\tau_s = \left[N_{Mn}\alpha^2 \frac{I_s}{\hbar^3}\frac{m^*}{4}[S(S+1) + \langle S_x^2 \rangle]\right]^{-1},$$

where $\alpha$ is the s-d exchange coupling constant, $m^*$ is the effective mass, $I_s$ is determined by the confinement of the quantum well. $\tau_s$ is inverse proportional to Mn concentration, $N_{Mn}$.

**Conclusion**

To study electron spin relaxation in p-type GaAs:Mn magnetic semiconductor, deep impurities of Mn and shallow impurities of donors and acceptors were calculated from the Hall measurements of hole concentrations in the valence bands. The temperature dependence of the effective degree of degeneracy of the magnetic acceptor filled with a hole, which arises due to the thermal filling of excited spin states, was taken into account. The concentrations calculated lead to the characteristics of the samples of p-type GaAs:Mn that demonstrate prolong electron spin relaxation times.

The observed electron spin relaxation time of 77 ns for the p-type quantum well GaAs:Mn is very long in comparison with earlier-reported results for p-type GaAs. Thus, the p-type 2D and 3D GaAs:Mn demonstrate the effectiveness for usage in different nanostructures with the determined concentrations of deep and shallow impurities. The composition of the p-type

GaAs:Mn is significant at the growth process, and should be accounted for when choosing the growth method as well as the growth parameters of the material.

**Acknowledgements**

The author thanks Prof. G. G. Zegrya for fruitful discussions during the work on the article.

| GaAs:Mn sample | $p(T_1)$, $10^{17}$ cm$^{-3}$ | $p(T_2)$, $10^{13}$ cm$^{-3}$ | $N_{Mn}$, $10^{17}$ cm$^{-3}$ | $N_d - N_a$, $10^{15}$ cm$^{-3}$ |
|---|---|---|---|---|
| 1 | 2.0 | 1.80 | 8.86 | 3.69 |
| 2 | 1.0 | 1.0 | 2.73 | 2.04 |
| 3 | 0.60 | 0.60 | 1.23 | 1.53 |
| 4 | 1.20 | 40.0 | 3.60 | -0.33 |
| 5 | 1.40 | 0.24 | 5.22 | 15.95 |
| 6 | 2.0 | 40.0 | 8.69 | -0.24 |
| 7 | 2.60 | 1.70 | 14.26 | 6.31 |
| 8 | 1.0 | 0.60 | 2.77 | 3.44 |
| 9 | 0.60 | 8.0 | 1.20 | 0.03 |
| 10 | 1.90 | 2.0 | 8.07 | 3.03 |
| 11 | 1.80 | 2.30 | 7.32 | 2.38 |
| 12 | 1.80 | 0.39 | 7.83 | 14.90 |
| 13 | 0.84 | 49.0 | 2.01 | -0.46 |
| 14 | 2.0 | 79.0 | 8.67 | -0.71 |
| 15 | 3.0 | 110.0 | 18.02 | -0.98 |
| 16 | 1.80 | 38.0 | 7.22 | -0.24 |
| 17 | 1.50 | 1.10 | 5.40 | 3.68 |
| 18 | 1.0 | 0.28 | 2.88 | 7.57 |
| 19 | 1.30 | 1.0 | 4.23 | 3.17 |
| 20 | 2.20 | 2.40 | 10.46 | 3.27 |
| 21 | 2.80 | 0.67 | 17.02 | 19.0 |
| 22 | 1.10 | 0.20 | 3.49 | 12.72 |
| 23 | 1.30 | 0.68 | 4.28 | 4.71 |
| 24 | 1.0 | 0.85 | 2.74 | 2.41 |

**Table 1**: Concentrations of valence-band holes $p$ and deep $N_{Mn}$ and shallow $N_d - N_a$ impurities in GaAs:Mn at $T_1 = 300K$ and $T_1 = 77K$.

**Figure Caption**

**Figure 1:**   The theoretical dependences $\tau_s(g)$ to fit the experimental data on relaxation times vs. the excitation power for 3D and 2D p-type GaAs:Mn.

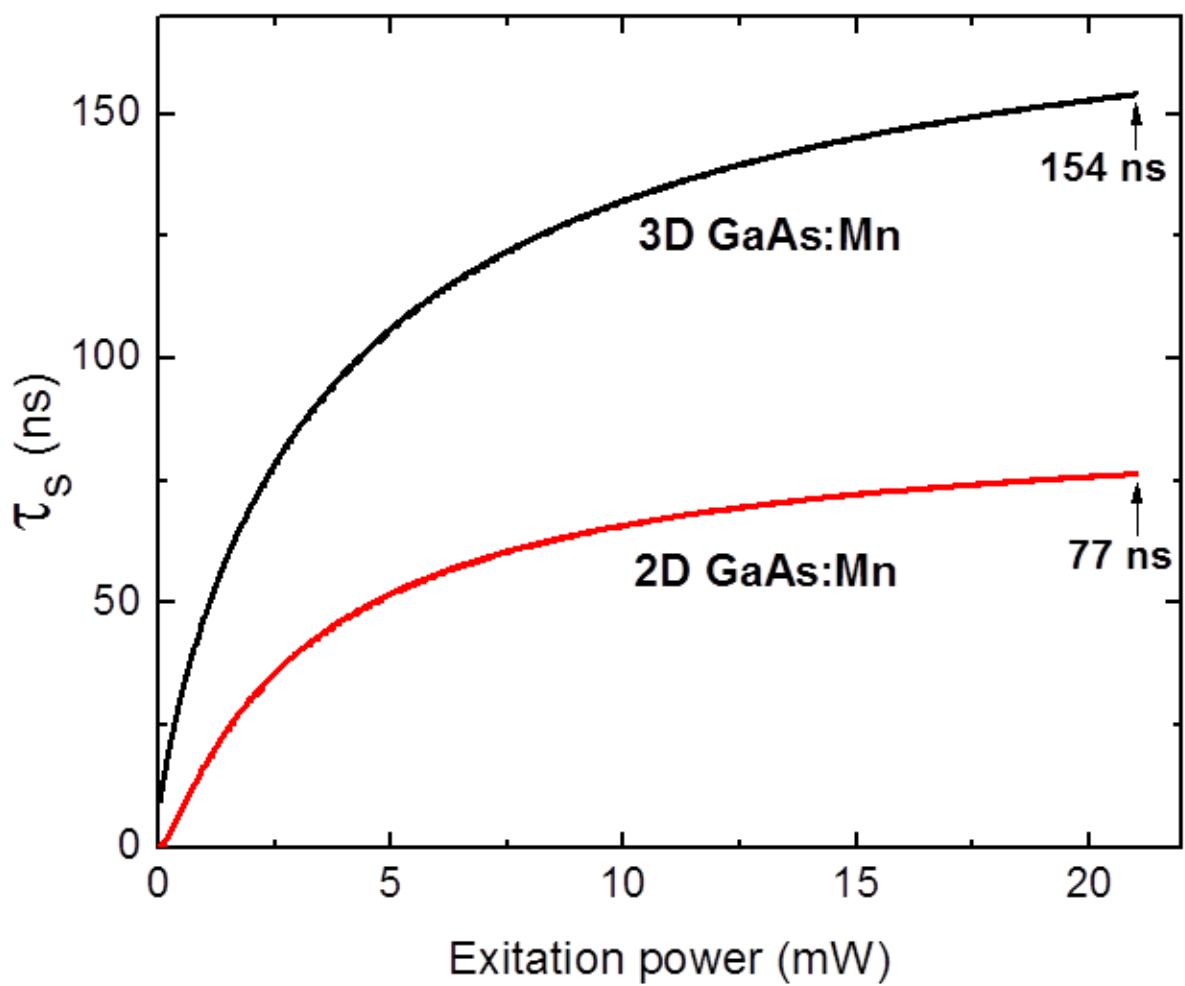